\documentclass{svjour3}
\usepackage{graphicx}%
\usepackage{url}%
\usepackage{cite}%
\usepackage{xcolor,colortbl}%
\usepackage{listings}%
\usepackage{comment}%
\usepackage{subcaption,ragged2e}%
\usepackage[misc,geometry]{ifsym}%
\usepackage[sort,compress]{natbib}
\renewcommand\fbox{\fcolorbox{black}{gray!30}}
\usepackage{multirow}
\usepackage{indentfirst}
\usepackage{array}
\definecolor{LightCyan}{rgb}{0.88,1,1}
\definecolor{blizzardblue}{rgb}{0.68, 0.85, 0.9}

\begin{document}

\title{An Empirical Study of Untangling Patterns of Two-Class Dependency Cycles}

\author{Qiong Feng    \and
        Shuwen Liu     \and
        Huan Ji      \and
        Xiaotian Ma  \and
        Peng Liang
}

\institute{Qiong Feng \and Shuwen Liu \and  Huan Ji \and Xiaotian Ma \at School of Computer Science, Nanjing University of Science and Technology, Nanjing, China \\ 
            \email{\{qiongfeng, hyggen, alex, xyzboom\}@njust.edu.cn}
            \and
            Peng Liang (\Letter) \at School of Computer Science, Wuhan University, Wuhan, China \\ 
            \email{liangp@whu.edu.cn}
}

\date{Received: date / Accepted: date}

\maketitle

\begin{abstract}
Dependency cycles pose a significant challenge to software quality and maintainability. However, there is limited understanding of how practitioners resolve dependency cycles in real-world scenarios. This paper presents an empirical study investigating the recurring patterns employed by software developers to resolve dependency cycles between two classes in practice. We analyzed the data from 38 open-source projects across different domains and manually inspected hundreds of cycle untangling cases. Our findings reveal that developers tend to employ five recurring patterns to address dependency cycles. The chosen patterns are not only determined by dependency relations between cyclic classes, but also highly related to their design context, i.e., how cyclic classes depend on or are depended by their neighbor classes. Through this empirical study, we also discovered three common counterintuitive solutions developers usually adopted during cycles' handling. These recurring patterns and common counterintuitive solutions observed in dependency cycles' practice can serve as a taxonomy to improve developers' awareness and also be used as learning materials for students in software engineering and inexperienced developers. Our results also suggest that, in addition to considering the internal structure of dependency cycles, automatic tools need to consider the design context of cycles to provide better support for refactoring dependency cycles.

\end{abstract}

\section{Introduction}
\label{sec:introduction}

In the ever-evolving world of software development, producing reliable, maintainable, and efficient code is very important. While developers strive to build quality software systems, there are certain pitfalls that can affect the quality and longevity of software systems, with cyclic dependency being a prominent one \citep{xiao:2021tse,oyetoyan:2013jss,mo:2019tse}. Cyclic dependency occurs when two or more modules, classes, or components in a system rely on each other directly or indirectly, forming a loop of dependencies \citep{lakos:1996large}. Cyclic dependency not only makes the code more challenging to understand, but also limits its reusability and testability. As these cycles proliferate in a software system, they create a complex, brittle structure that is resistant to change and more prone to errors \citep{xiao:2021tse,mo:2019tse}.

While several automated tools can help detect and visualize dependency cycles such as Designite\footnote{\url{https://www.designite-tools.com}}, Structure101\footnote{\url{https://structure101.org}}, and SonarQube\footnote{\url{https://www.sonarqube.org}}, there is limited understanding of how practitioners resolve or refactor these cycles in real-world scenarios. A recent study \citep{feng:2023jsep} shows that untangling strategies of a dependency cycle are highly correlated with its topological structure and the dependency relations inside the cycle. For example, a previously circle-shaped dependency cycle was broken into a circle-shaped dependency cycle of a smaller size. Though this study reveals dependency cycles evolution patterns in terms of structure and dependency types, some of important details in the untangling process of dependency cycles still remain unclear. For example, we do not know when a dependency relation is removed to resolve a cycle, whether or how the system maintains its original functionality. We also do not know the specific challenges that practitioners face while resolving dependency cycles.


To better understand dependency cycle' resolution and challenges during the process, we conduct a large-scale empirical study of dependency cycles' fix in commits to understand whether and how the original functionalities get maintained when a dependency relation in dependency cycle is removed. Particularly we focus on dependency cycles' resolution between two classes. There are two reasons. First, a previous study found that most of dependency cycles' resolutions are not a one-shot process \citep{feng:2023jsep}. Instead, it involves multiple steps: first to a less complex cycle (which sometimes involves two classes' resolution), then to a two-classes cycle, and finally get fully untangled. Thus, two-class cycles' resolution is usually the last and also an critical step in the whole untangling process. Second, two-class cycles' resolutions are the most frequent cases in code repositories of multiple open-source projects \citep{feng:2023jsep}. This finding shows that developers have spent a significant effort on this type of cases. A study of two-class resolution can reveal the specific challenges that practitioners face while resolving such dependency cycles and also provide valuable insights into the whole untangling process.

To this end, we designed an approach to identify recurring untangling patterns employed by software developers to resolve dependency cycles between two classes. Using the data from 38 open-source projects, we first extracted source code and structural dependency graphs before and after each commit, and we detected dependency cycles before and after each commit. Then for each commit, we applied an algorithm defined in \cite{feng:2023jsep} to detect whether the commit involves the fix of dependency cycles. In this way, we detected 587 successful and 69 unsuccessful untangling cases in these 38 open-source projects. Next, we manually located the code snippet which causes the dependency cycles and the code changes to fix it. At last, we followed a rigorous process to summarize dependency cycles' fix patterns. Our investigations reveal the \textbf{following results}:

First, developers tend to use five recurring fix patterns in two-class dependency cycles' successful untangling. 91.3\% (536/587) of two-class dependency cycles' untangling cases are resolved by these patterns, including ``Remove Unused/Deprecated Code'' (40.0\%), ``Move Code From One Cyclic Class To a Third Class''(23.4\%), ``Move Code Between Two Cyclic Classes'' (14.8\%),  ``Shorten Call Chain''(12.1\%), and ``Leverage Built-In Feature'' (1.0\%), which will be described later in Section~\ref{sec:results}. Since these recurring patterns can be observed in different projects with various domains, we are confident that a taxonomy of these patterns can help developers understand dependency cycles better and thus tackle them more efficiently.

Second, not only is the chosen untangling pattern related with the internal dependency relations inside a dependency cycle, but also highly affected by the cycle's design context, i.e., how cyclic classes depend on or are depended by its neighbor classes. We found that the internal dependency relations inside a dependency cycle can be used to predict its untangling pattern. Moreover, for patterns which involve a third class's participation, combining a cycle's design context with its internal structure can help better predict its untangling pattern.

Third, among all cases involving addressing dependency cycles, we also found that 10.5\% (69/(69+587)) of cases were not successfully untangled. It means that, in around every 10 dependency cycles' untangling cases, there exists 1 dependency cycle which is not properly handled. These counterintuitive cases can be classified into three categories. We believe that the awareness of these categories can help developers avoid similar problems in future development of untangling dependency cycles.

Overall, this paper reveals the recurring fix patterns and common challenges when developers are addressing two-class dependency cycles. This paper also proves that the design characteristics inside and outside a dependency cycle can determine the chosen untangling pattern.

The rest of the paper is organized as follows. Section~\ref{sec:approach} presents the general approach. Section~\ref{sec:rq} introduces the open-source projects we used in this study and the corresponding research questions. Section~\ref{sec:results} and Section~\ref{sec:diss} discuss the results and how these results can benefit developers in addressing dependency cycles. Section~\ref{sec:related} lists related work and Section~\ref{sec:conclude} concludes this paper with future directions.
\section{Approach}
\label{sec:approach}

\begin{figure*}[!thbp]
	\centering
	\includegraphics[width=\linewidth,trim={2cm 6cm 11cm 1cm}]{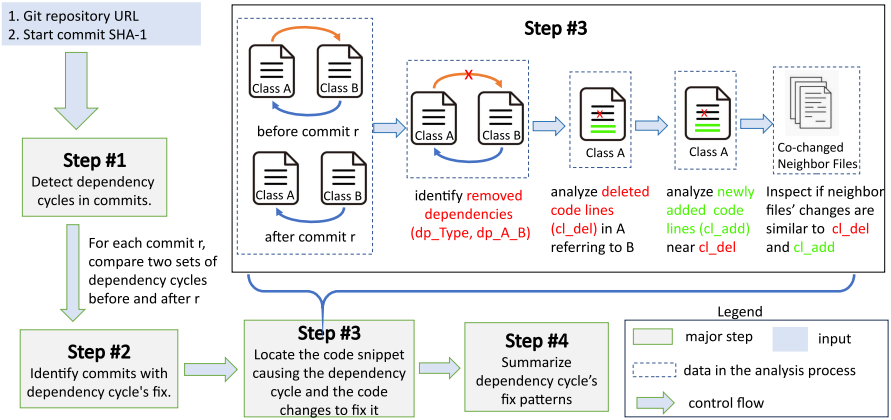}
        \centering
	\caption{Overview of Pattern Identification Approach}
	\label{fig:approach}
\end{figure*}


Numerous research have explored code change patterns in various scenarios such as program fixes, regression repairs, and API misuse \citep{meng:2013lase, tan:2015relifix,kim:2013automatic,koyuncu:2020fixminer, liu:2021identifying}. These studies commonly employ a practice that involves locating specific commits, identifying code changes in those commits, analyzing the relationship between added and removed code snippets, and summarizing the resulting change patterns. Notably, these studies often identify patterns through manual inspections of code changes and subsequently verify the recurrence of these patterns across multiple projects \citep{liu:2021identifying, meng:2013lase}. In our work, we also adopted this practice to locate fix commits for dependency cycles and examined the code changes within those commits. Resolving dependency cycles typically entails modifications in multiple code locations. Therefore, our analysis goes beyond analyzing added and removed code snippets solely within the cyclic classes. We also traced the calling chain in the modified code of these classes and explored whether similar code changes exist in other source files. This consideration is important because code changes can propagate through both structural and semantic relations. By tracing the calling chain of added or removed code snippets and grouping similar code changes, we have observed six recurring patterns, which will be further discussed in the Experiment Results section (Section~\ref{sec:results}). The overall approach for detecting recurring patterns in the resolution of dependency cycles comprises four steps, which are illustrated in Figure~\ref{fig:approach}.


\textbf{Step \#1:} Detect dependency cycle in commits. To identify dependency cycles within each commit, we first extract the source code by executing ``\textsc{git show changed\_filenames}''. Then we utilize an open source reverse engineering tool \textit{Depends}\footnote{\url{https://github.com/multilang-depends/depends}} to extract dependency relationships between source files before and after each commit. \textit{Depends} can parse entities in Java source code and identify 13 unique kinds of dependency relations between entities, as shown in Table~\ref{tbl:dp_types}. Dependency relations among methods and variables are aggregated at the class level. Subsequently, we represent classes (nodes) and their structural dependencies (edges) as a directed graph. Utilizing the two graphs produced before and after each commit, we apply Kosaraju's algorithm \citep{sharir:1981strong} to detect strongly connected components (SCCs) within each graph. By definition, in an SCC, a node (class A) directly or indirectly relies on another node (class B), and likewise, class B directly or indirectly relies on class A. We opt to identify an SCC as a dependency cycle since a modification in one class is highly likely to propagate to another class within the same SCC.

\begin{table}[th]
	\centering
        \caption{Dependency Types Extracted by \textit{Depends}}
	\label{tbl:dp_types}
\begin{tabular}{|c|l|}
\hline
\textbf{Dependency Type} & \textbf{Description}                            \\ \hline  
{\sffamily call}     & a statement in a method invokes a method                        \\ \hline
{\sffamily cast}     & a statement in a method casts another type to a variable        \\ \hline
{\sffamily contain}  & a file contain classes, enums or other types                    \\ \hline
{\sffamily create}   & a statement in a method creates an instance of a type           \\ \hline
{\sffamily extend}   & a class extends a parent class                                  \\ \hline
{\sffamily implement}& a class implements an interface                                 \\ \hline
{\sffamily import}   & a file imports another class, enum, static method or attribute  \\ \hline
{\sffamily parameter}& a method use another type as its parameter                      \\ \hline
{\sffamily return}   & a method returns another type                                   \\ \hline
{\sffamily set}      & a statement in a method sets a variable's value                 \\ \hline
{\sffamily typed}    & a variable is initiated from a class or other types             \\ \hline
{\sffamily throw}    & a method throws an exception                                    \\ \hline
{\sffamily use}      & a method uses a local variable or parameter in its scope        \\ \hline
\end{tabular}
\end{table}

\textbf{Step \#2:} Identify commits with dependency cycle's fix. Upon identifying dependency cycles both before and after each commit, we compare the two sets of dependency cycles from these respective states. We adopt the method to identify dependency cycle' fix in this paper~\citep{feng:2023jsep}. If two classes formed a dependency cycle, but are no longer in the same dependency cycle after a commit, we consider it as a candidate for a dependency cycle fix. There are three possibilities: at least one class gets deleted, both classes exist independently (not in any dependency cycle), or these two classes get untangled and at least one class joins a dependency cycle with other classes in the system. After we identify the commit with dependency cycles' fix, we execute  ``\textsc{git diff changed\_filenames}'' to extract code changes associated with each class involved in the commit. This diff information contains details related to how the dependency cycle was fixed, which will be the input for the next step.

\textbf{Step \#3:} Locate the code snippet causing dependency cycle and the code changes to fix the cycle. This is the most time-consuming step in our approach as it involves several minor steps and also relies on manual inspections. We establish a protocol for this process. First, since the dependency cycle is disentangled, we identify the removed dependency relationships between two cyclic classes, such as when class A eliminates its call to a method of class B, by comparing the dependency graph before and after the commit. We mark the removed dependency relation's type. Next, we analyze deleted code lines in class A and search for the deleted code snippet referring to class B. We validate the deleted code snippet by matching with removed dependency types and its location. For example, if the dependency graphs indicates class A removes its ``call'' to class B, we check whether the deleted code in class A contains \textsc{b.func()} to validate the deleted code line is actually the line for the dependency change. Subsequently, we check whether any new code snippets are added in class A near the location of the deleted code line. If present, we examine if the added code calls other classes or types and manually inspect whether the added code provide similar features as the deleted code, such as containing the similar method signature or code structure. This step required the dedicated efforts of four students with 4 years of Java development experience, and one senior researcher with 14 years of Java Development experience. We followed the practice outlined in \cite{campbell:2013coding}, allowing the senior researcher to establish the procedure for identifying code changes and training the students with examples. Subsequently, the four students were divided into two groups, with two students in each group collaborating to identify code changes. We then cross-examined the results from both groups to reconcile any disagreements and reach a consensus. For instance, one disagreement occurred when one group thought that added code changes were to replace the same  feature of deleted code lines, while the other group believed that the purpose of the added code lines was not to provide the same feature. Another disagreement occurred when one group thought that the removed code lines were not responsible for the untangling of the dependency cycle, but the other group disagreed. In the event of such disagreements, the senior researcher would review the specific case and discuss it with the students until a consensus was reached. Beside the two cyclic classes, we also track changed code snippet of other source files in the same commit and analyze the type calling in the deleted and added code.


\begin{figure*}[!thbp]
	\centering
	\includegraphics[width=\linewidth]{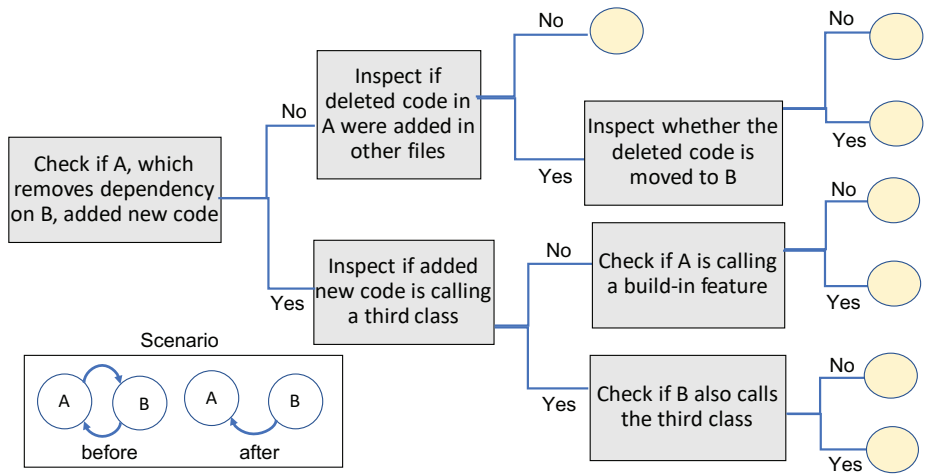}
        \centering
	\caption{Process of Identifying Untangling Patterns}
	\label{fig:identifyPattern}
\end{figure*}

\textbf{Step \#4:} Summarize dependency cycles' fix patterns. Through Step \#3, we obtain code changes in dependency cycle and their propagation to other neighbor files for fixing dependency cycles. We then analyze numerous cases across different commits and projects to identify potential recurring fix patterns. To do this, we design a decision tree strategy to study fix patterns of dependency cycles, as shown in Figure~\ref{fig:identifyPattern}. The rationale behind this strategy is to investigate whether a specific feature remains intact during the resolution of a dependency cycle and to determine the scope of the impact if the feature is maintained through code changes. In particular, we examine the extent to which code changes affect both the cyclic files and neighboring files. By doing so, we gain insights into how the dependency cycles' fix influence the functionality and dependencies within the system.

Suppose that there is a cycle involving two source classes, A and B, and this cycle is being untangled in a commit where class A no longer calls class B. In order to analyze this situation, we first locate the old code snippet in class A that was causing the dependency cycle and check if any new code snippets were added in its context. In case there are no new code snippets in class A, we examine whether the deleted code snippet can be found in other files. If the deleted code snippet is not found anywhere, it suggests that the feature in class A may have been simply removed. However, if the deleted code snippet is found in another location, we investigate whether it has been relocated to the other cyclic class B, or to a third file to determine the change scope. 

On the other hand, if new code snippets were indeed added in the source file of class A, we examine whether these new additions provide similar features as the deleted code snippets. If similarities are found, we further inspect whether the types or classes referenced in the new code snippets are consistent with the original ones and trace up the referenced types or classes. This step can help us identify whether code changes are limited to only one cyclic class, or both cyclic classes, or affecting more neighboring classes. Depending on the change scope and the referenced types, we categorize the fix into different categories. For example, if the new code does not call any other type or class, we check whether it utilizes built-in features, and classify the fix accordingly.


By following these steps, we discover recurring fixes that lead to the identification of several common patterns for fixing dependency cycles.

\section{Research Questions}
\label{sec:rq}

\subsection{Subjects}

The objective of our study is to explore the recurring patterns of dependency cycle resolution in various projects. To achieve this, we first leverage the dataset from our previous study of dependency cycle~\citep{feng:2023jsep}, which includes 12 Apache projects and 6 other projects from different domains. To ensure the generality of observed patterns, we also crawled dependency cycle resolution cases from other active communities: Google, Eclipse, Facebook, and Alibaba. These communities are well-known, and the projects they open-sourced are generally considered reliable. We utilized the GitHub API and conducted a search using the criteria `\textsc{stars:$>$100 language:Java forks:$>$100}' to identify popular and active projects within these communities. Subsequently, we selected the top 5 most contributor projects in each community to ensure that we capture patterns from different developers. This process resulted in a total of 38 open-source projects. Detailed information about all these projects can be found in our replication package~\citep{Replication}.

Due to space limitations, we select 3 projects with most contributors from each community and present the demographic information of these 18 projects in Table~\ref{tbl:subjects}. The first column, labeled \textbf{Cmty}, indicates the community to which the projects belong. Columns 2-5 display the project's name, number of Java files, commits, and contributors. The last column indicates the project's domain. These projects vary in size from 52 to 14,684 Java files, with commit counts ranging from 266 to 37,028, and development histories spanning 4 to 21 years. By investigating the details of changes made during the resolution of dependency cycles in these projects, we aim to gain a deeper understanding of how dependency cycles are resolved in practice and the challenges associated with this process.

\begin{table}[th]
	\caption{{Demographic Information about the Study Subjects}}
	\centering
	\label{tbl:subjects}
\begin{tabular}{|c|c|r|r|r|c|}
\hline

\textbf{Cmty} &\textbf{Proj} & \textbf{\#Java} & \textbf{\#Cmt} & \textbf{\#Ctr} & \textbf{Domain} \\ \hline  
\multirow{3}{*}{Apache} &Avro   & 661 & 2435 & 296 & Data Process \\ \cline{2-6}  
& Cassandra   & 2751 & 25234 & 274 & Database System \\ \cline{2-6}  
& HBase  & 4233 & 17621 & 263 & Data Process \\ \hline   \hline 
\multirow{3}{*}{Google} & guava & 3199 & 6165 & 299 & Utility Core Library \\ \cline{2-6}  
& closure-compiler   & 1302 & 18774 & 256 & Complier  \\ \cline{2-6}   
& ExoPlayer  & 1638 & 18864 & 247 & Media Player \\ \hline   \hline  
\multirow{3}{*}{Eclipse} & jetty.project  & 3365 & 25066 & 183 & Web Server \\ \cline{2-6}  
& jkube  & 1187 & 1650 & 162 & Deployment \\ \cline{2-6}  
& eclipse-collections   & 2664 & 1650 & 162 & Data Structure Framework \\  \hline  
\multirow{3}{*}{Facebook} & buck   & 6996 & 22696 & 289 & Build System \\ \cline{2-6}  
& fresco   & 808 & 3649 & 218 & Image Loading and Display \\ \cline{2-6}  
& litho   & 1573 & 16447 & 205 & UI Generator \\ \hline  \hline 
\multirow{3}{*}{Alibaba} & nacos  & 2119 & 4703 & 330 & Service Data Manager \\ \cline{2-6}  
& druid & 5138 & 6874 & 205 & JDBC component library \\ \cline{2-6}  
& Sentinel  & 1280 & 815 & 184 & Service Manager  \\ \hline  \hline  
\multirow{3}{*}{Others} & Javaparser  & 1751 & 8029 & 183 & Code Analyzer  \\ \cline{2-6}  
& Stripe-java  & 787 & 2344 & 143 & Payment System \\ \cline{2-6}  
& Spoon & 2055 & 3882 & 119 & Code Analyzer \\ \hline
\end{tabular}
\end{table}

\subsection{Research Questions}
We investigated the following Research Questions (RQs) to gain insight into the methods used for resolving dependency cycles in practice and the frequent errors that arise during this process.

\textbf{RQ1}. \textit{Are there any recurring patterns developers employ to resolve dependency cycles?} Recognizing common patterns for addressing dependency cycles is crucial for automating their resolution. If we can identify the recurring patterns, we can establish a taxonomy that developers can consult when attempting to resolve dependency cycles.

\textbf{RQ2}. \textit{Does each recurring pattern exhibit specific design characteristics in terms of dependency types}? If dependency cycles with particular design characteristics tend to be resolved by certain recurring patterns, we can recommend developers to choose a specific pattern or even automate the resolution process. On the other hand, if dependency cycles with similar design characteristics are resolved by different types of recurring patterns, it indicates that the patterns are not related to design characteristics.

\textbf{RQ3}. \textit{Are there any counterintuitive solutions when developers try to resolve dependency cycles, and why?} According to the work of \cite{lu:2016icse} and \cite{feng:2019ase}, new anti-patterns could be introduced when existing anti-patterns get resolved. However, previous research has not examined how and why new anti-patterns are introduced. By addressing this RQ, we plan to explore the challenges developers face and the counterintuitive solutions developers make while resolving dependency cycles and discuss the strategies to avoid them.



\section{Experiment Results}
\label{sec:results}

\subsection{Recurring Untangling Patterns}
\label{patterns}

\begin{figure*}
\captionsetup[subfigure]{justification=Centering}
\centering
\begin{subfigure}[t]{0.31\textwidth}
    \includegraphics[width=\textwidth]{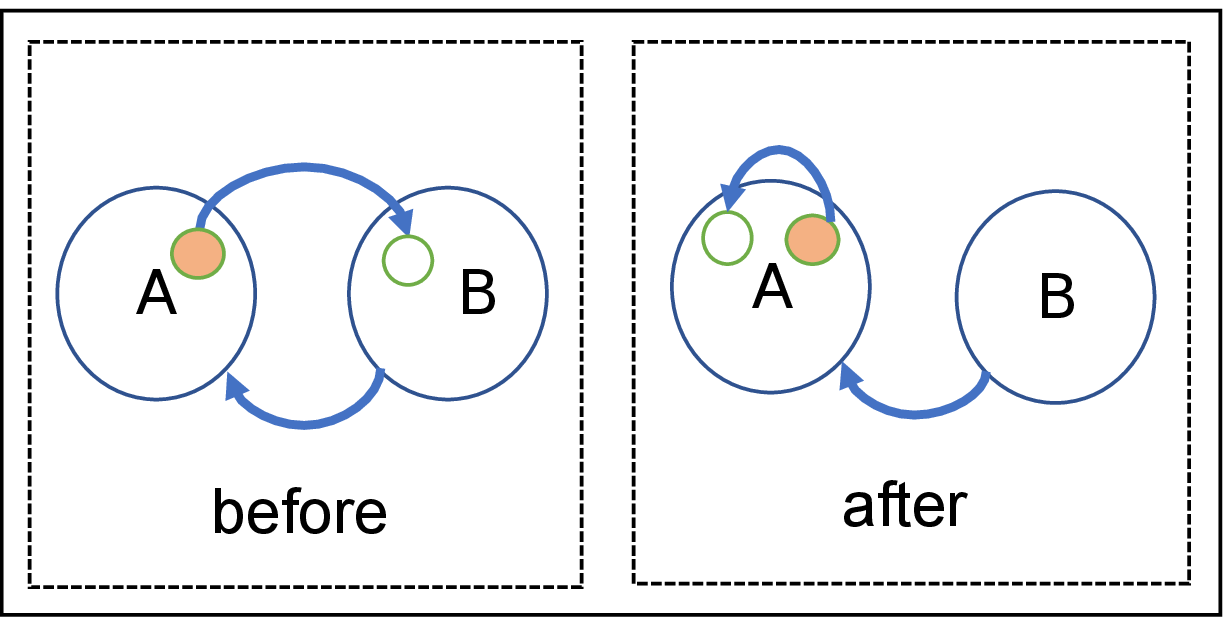}
    \caption{Move Code between Two Cyclic Classes}
    \label{fig:pattern1}
\end{subfigure}\hspace{\fill} 
\begin{subfigure}[t]{0.31\textwidth}
    \includegraphics[width=\linewidth]{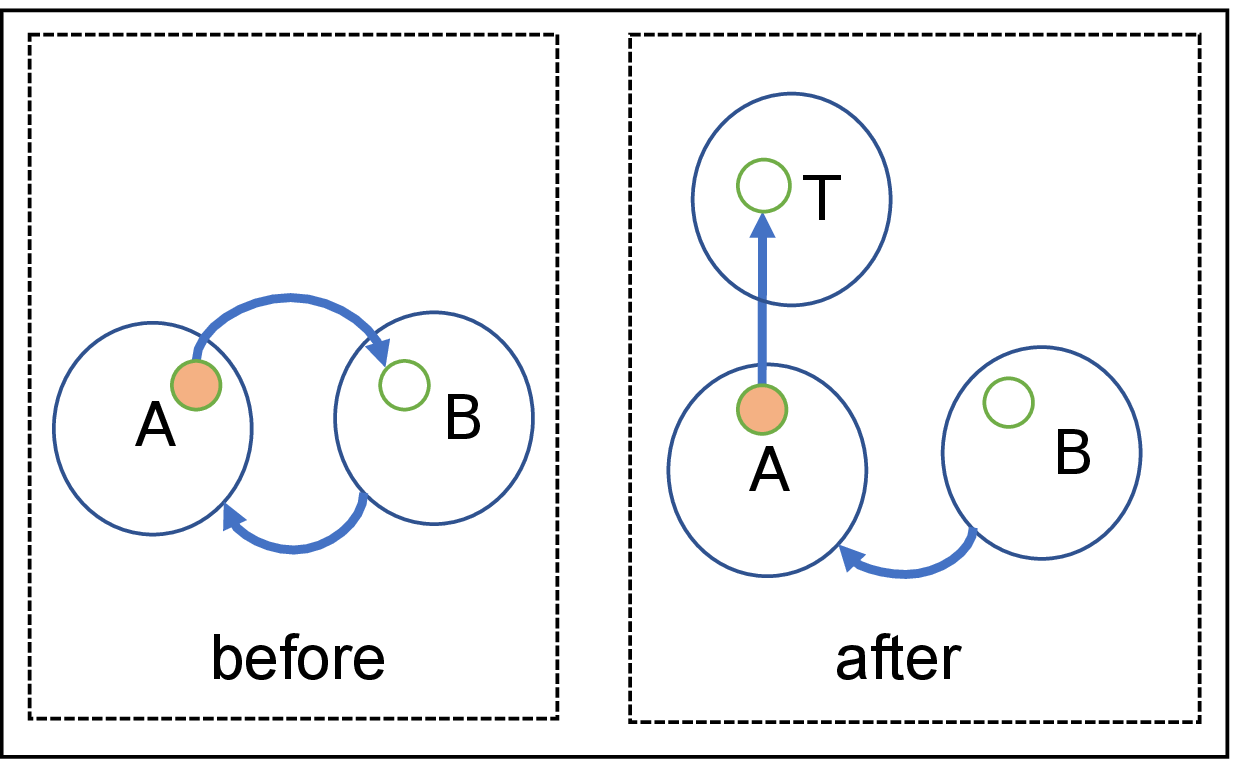}
    \caption{Move Code from One Cyclic Class to a Third Class}
    \label{fig:pattern2}
\end{subfigure}\hspace{\fill} 
\begin{subfigure}[t]{0.31\textwidth}
    \includegraphics[width=\linewidth]{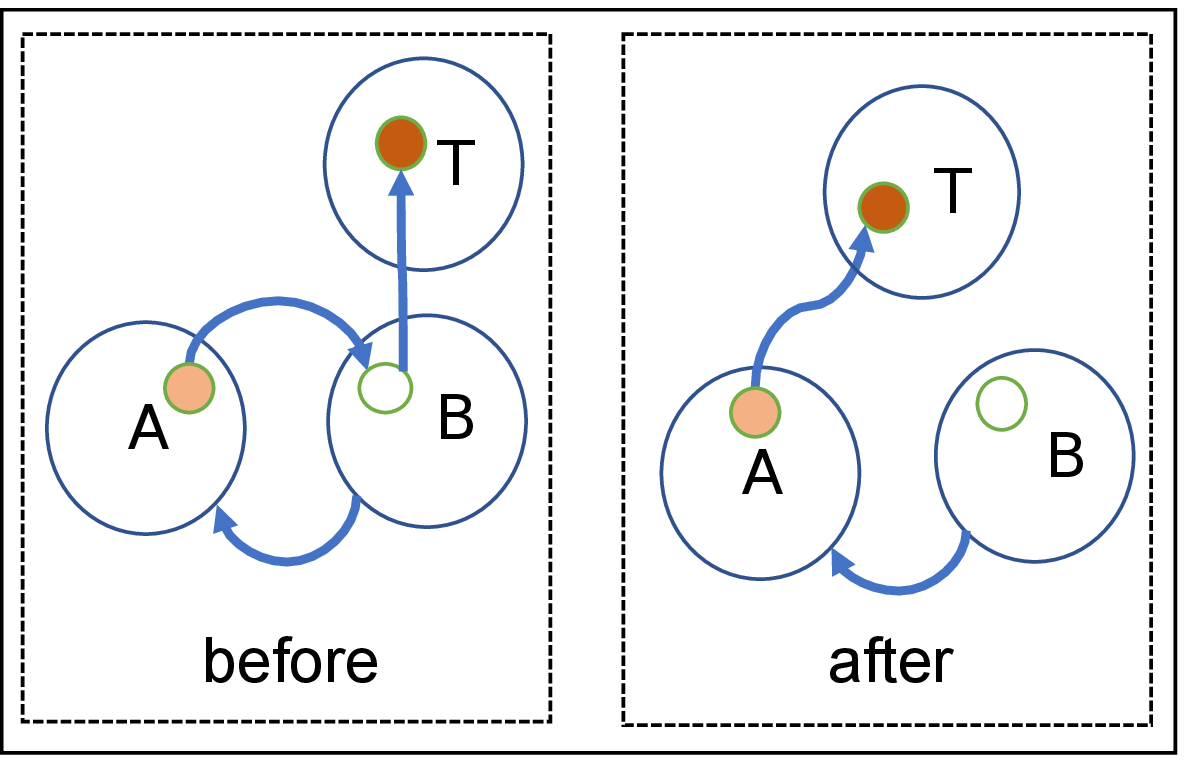}
    \caption{Shorten Call Chain}
    \label{fig:pattern3}
\end{subfigure}

\bigskip 
\begin{subfigure}[t]{0.33\textwidth}
    \includegraphics[width=\linewidth]{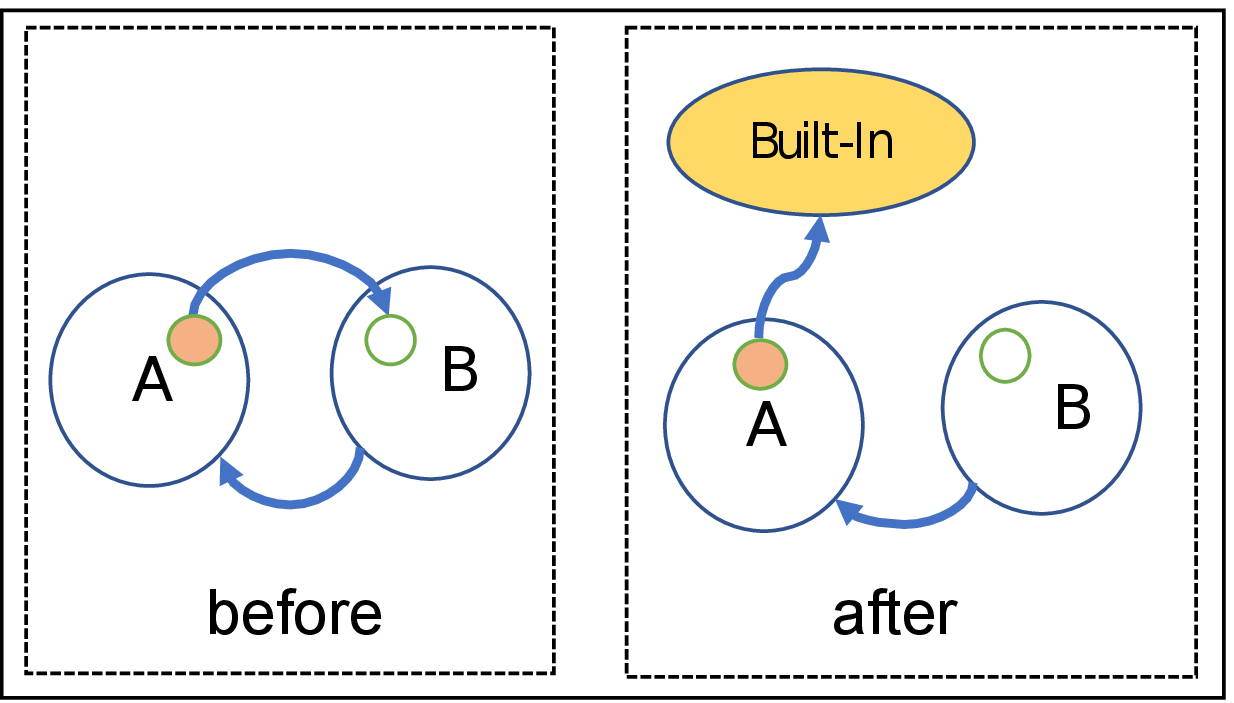}
    \caption{Leverage Built-In Feature}
    \label{fig:pattern4}
\end{subfigure} 
\begin{subfigure}[t]{0.33\textwidth}
    \includegraphics[width=\linewidth]{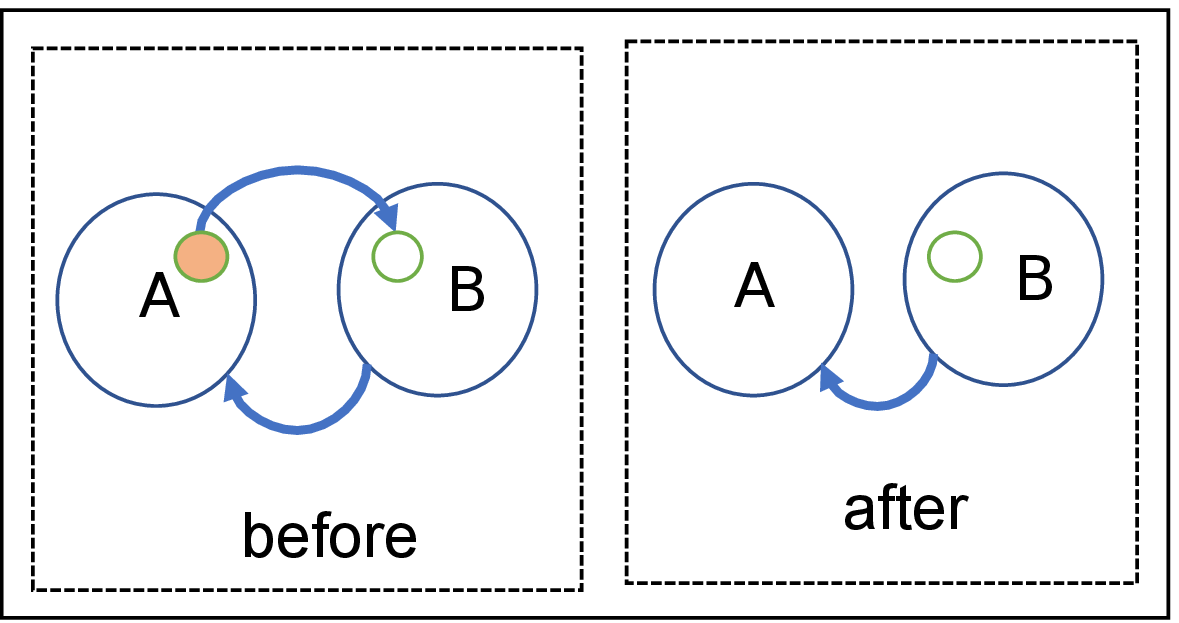}
    \caption{Remove Unused Code}
    \label{fig:pattern5}
\end{subfigure}

\caption{Recurring Patterns in Dependency Cycles' Untangling}
\label{fig:recurrpattern}
\end{figure*}

By employing the analysis strategy outlined in Section~\ref{sec:approach}, we discovered 587 dependency cycles' fix from 38 open-source projects and analyzed how they were resolved by contributors in practice. Among these cases, we discovered six recurring patterns, as shown in Table~\ref{tbl:count}. From this table, we can see ``Remove Unused/Deprecated Code'' and ``Move Code From One Cyclic Class To a Third Class'' is the two most frequently observed patterns while ``Leverage Built-In Feature'' is the least common pattern. In this section, we discuss these recurring patterns, which are illustrated in Figure~\ref{fig:recurrpattern}.

\begin{table}[th]
	\caption{The Statistics of Untangling Patterns}
	\centering
	\label{tbl:count}
\begin{tabular}{|l|r|r|}
\hline
\textbf{Untangling Patterns} & \textbf{Count} & \textbf{Percentage} \\ \hline 
Remove Unused/Deprecated Code & 235 & 40.0\%\\ \hline 
Mode Code From One Cyclic File to a Third Class & 137 & 23.4\% \\ \hline 
Move Code Between Two Cyclic Files & 87 & 14.8\% \\ \hline 
Shorten Call Chain & 71 & 12.1\% \\ \hline 
Complex & 51 & 8.7\%\\ \hline 
Leverage Built-In Feature & 6 & 1.0\%\\ \hline 
\textbf{Total} & 587 & 100\%\\ \hline 
\end{tabular}
\end{table}

\subsubsection{Move Code Between Two Cyclic Classes}
As shown in Figure~\ref{fig:pattern1}, to resolve the cycle of class A and class B, the code in class B, which was referenced by class A, was moved from class B to A. Thus, class A no longer depends on class B. Instead, class A relies on the relocated code in itself. This case is similar to the feature envy code smell~\citep{fowler:1999book}. According to the definition of feature envy, when a method, attribute, or inner class in class B is more closely associated with class A than its own class, it should be moved to class A, eliminating the need for class A to call class B. 

\begin{figure}[!thbp]
	\centering
	\includegraphics[angle=90,width=\linewidth,trim={7cm 2cm 6cm 2cm}]{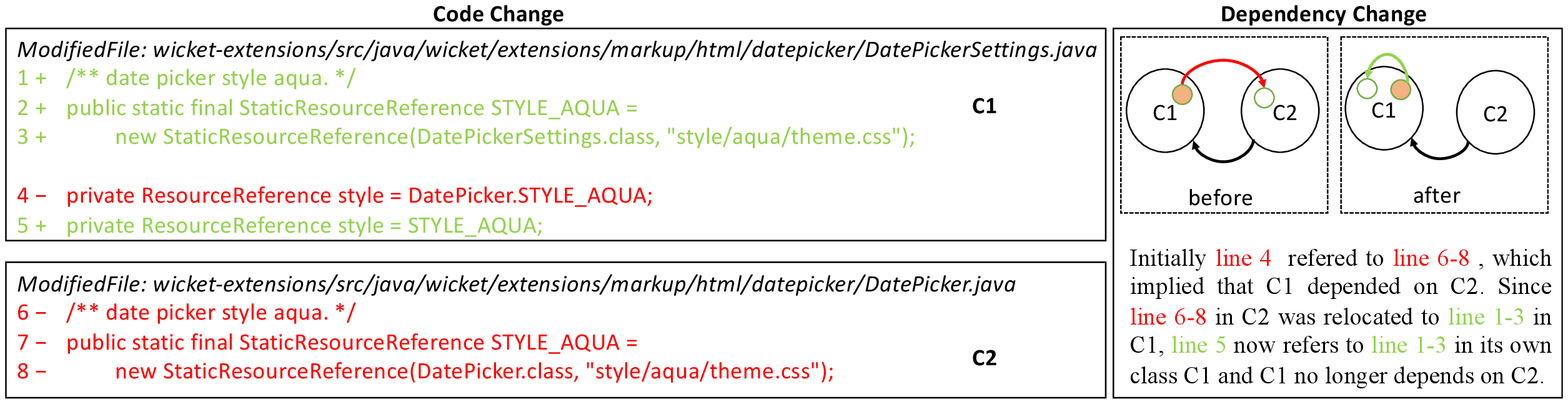}
	\caption{Move Code Between Two Cyclic Classes}
	\label{fig:moveBetweenAB}
 
\end{figure}

Figure~\ref{fig:moveBetweenAB} presents commit \#dec227e\footnote{\url{https://github.com/apache/wicket/commit/dec227e}} of the Wicket project. Class \textit{DatePicker} (denoted as C2) and \textit{DatePickerSettings} (denoted as C1) formed a cycle in which class C1 also uses the constant \textit{STYLE\_AQUA} of class C2 as shown in Line 4. To address this dependency cycle, the constant in class C2 were moved to class C1 and Line 6-8 in C2 was relocated to Line 1-3 in class C1. This eliminated the need for class C1 to call class C2. Furthermore, the code Line 4 \textit{DatePicker.STYLE\_AQUA} was replaced by~\textit{STYLE\_AQUA} in Line 5 after resolving the cycle.

\subsubsection{Move Code From One Cyclic Class To a Third Class} 

This scenario is different from the feature envy code smell. Although one cycle class relies on another cycle class, the method in the latter class is moved to a third class. It is usually designed for better scalability. As shown in Figure~\ref{fig:pattern2}, class A calls a method of class B. To untangle this dependency cycle, class A switches its dependency of class B to a third class by calling a similar method in the third class. 

\begin{figure}[!thbp]
	\centering
	\includegraphics[angle=90,width=\linewidth,trim={3cm 1cm 4cm 1cm}]{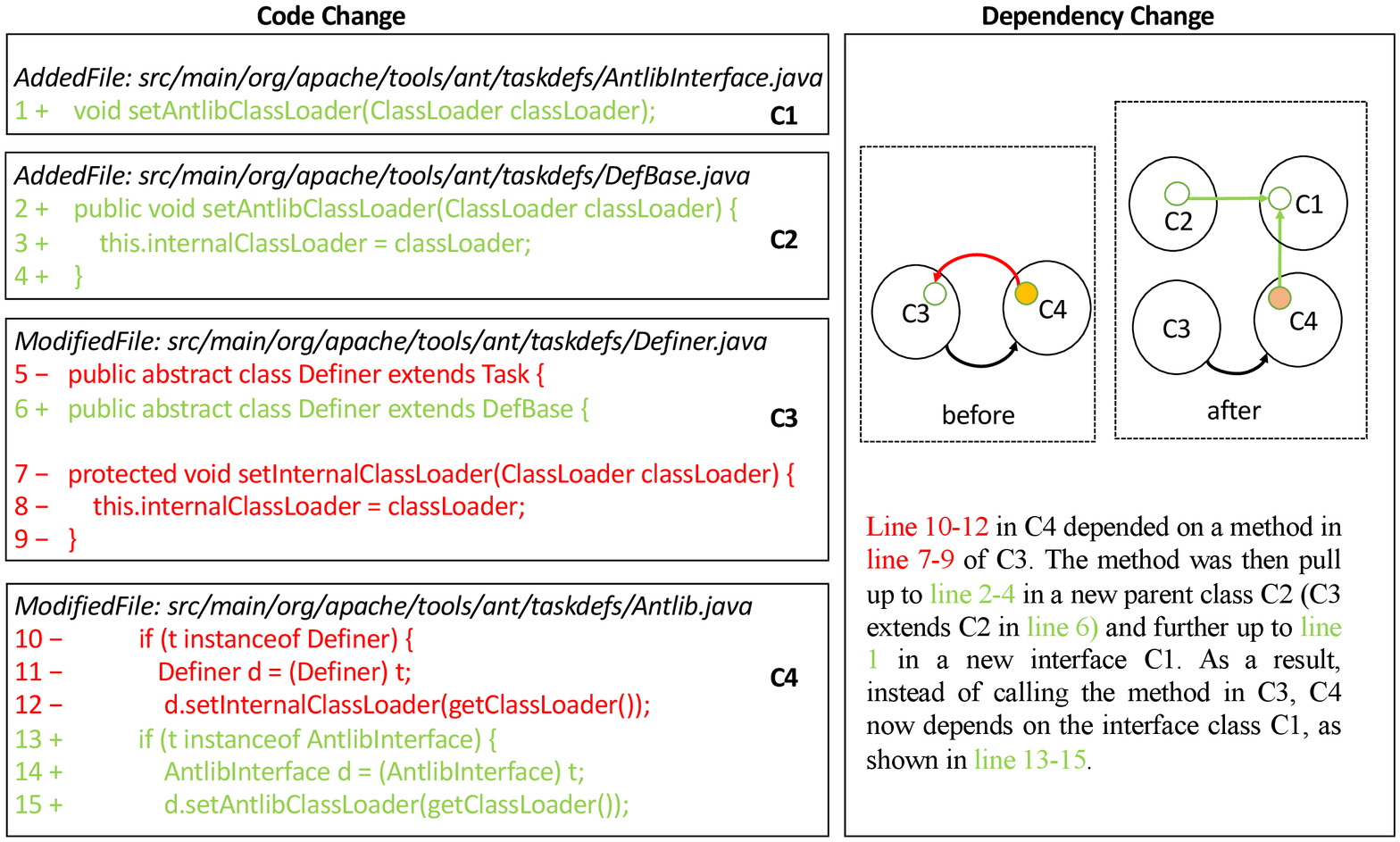}
	\caption{Move Code From One Cyclic Class To a Third Class}
	\label{fig:move2third}
\end{figure}

Figure~\ref{fig:move2third} presents an example of this case Commit \#ca5799f\footnote{\url{https://github.com/apache/ant/commit/ca5799f}} of the Ant project. Before this commit, class \textit{Antlib} (denoted as C4) and class \textit{Definer} (denoted as C3) formed a cyclic dependency relationship and class C4 called a method~\textit{setInternalClassLoader} of class C3, as shown in Line 10-12. During the commit, the method~\textit{setInternalClassLoader} in Line 7-9 of class C3 was removed and abstracted into a method~\textit{setAntlibClassLoader} of 2 new classes: \textit{AntlibInterface} (denoted as C1) and \textit{DefBase}  (denoted as C2). To accompany with this change, class C4 removed the call to the original \textit{setInternalClassLoader} method of class C3. Instead, it calls the \emph{abstracted} method in class C1. As shown in \emph{dependency change} of Figure~\ref{fig:move2third}, the red curve of C3 $\rightarrow$ C4 was removed and class C4 added its dependency on class C1, marked as the green curve. This case demonstrates that the cyclic dependency between two classes can be alleviated by shifting responsibilities to a third class. It is worth noting that the third class and removed code class likely have a parent-child relationship.

\subsubsection{Shorten Call Chain} 

As shown in Figure~\ref{fig:pattern3}, in the dependency cycles of class A and B, class A calls a method in class B, but the method of class B calls another third class. To resolve this cycle, class A does not call the method of class B but calls the third class directly, breaking the dependency cycle between class A and B, and further shortening a dependency chain of three classes to a chain of two classes.  

\begin{figure}[!thbp]
	\centering
	\includegraphics[angle=90,width=\linewidth,trim={5cm 1cm 5cm 1cm}]{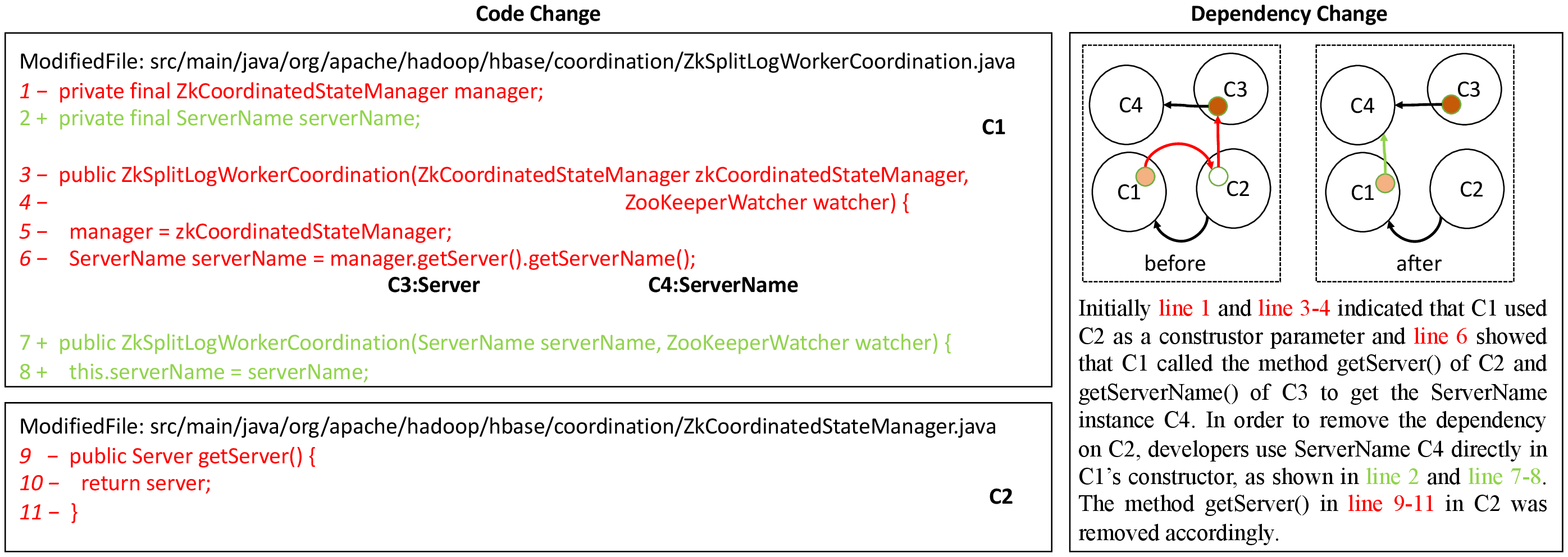}
	\caption{Shorten Call Chain}
	\label{fig:shortencallchain}
\end{figure}

Commit \#dd70cc3\footnote{\url{https://github.com/apache/hbase/commit/dd70cc3}} in the HBase project demonstrates an example of such cases. As shown in Figure~\ref{fig:shortencallchain}, initially class \textit{ZkSplitLogWorkerCoordination} (denoted as C1) calls  class \textit{ZkCoordinatedStateManager}'s (denoted as C2) method chain (C2's method \textit{getServer} and Server---C3's  method \textit{getServerName}) to get the class \textit{ServerName} in Line 6. To resolve this cycle, developers use class \textit{ServerName} (denoted as C4) in class \textit{ZkCoordinatedStateManager}'s constructor in Line 7 and directly call class \textit{ServerName} instead of going through the method calling chain of class C2 and C3. At the same time, class C2 removes \textit{ServerName} from its constructor in Line 9-11. As shown in \emph{Dependency Change} in Figure~\ref{fig:shortencallchain}, the red curve of C1$\rightarrow$C2 and C2$\rightarrow$C3 was eliminated and the green curve of C1$\rightarrow$C4 is added. In this case, not only is a cycle resolved, but a calling chain involving three classes is also shortened.

\subsubsection{Leverage Built-In Feature}

As shown in Figure~\ref{fig:pattern4}, this pattern uses class's build-in features instead of calling one cyclic class's method. In this scenario, instead of calling a method from one cycle class, the fix utilizes built-in features of the Java library to achieve similar functionality. 

\begin{figure}[!thbp]
	\centering
	\includegraphics[angle=90,width=\linewidth,trim={3cm 1cm 3cm 1cm}]{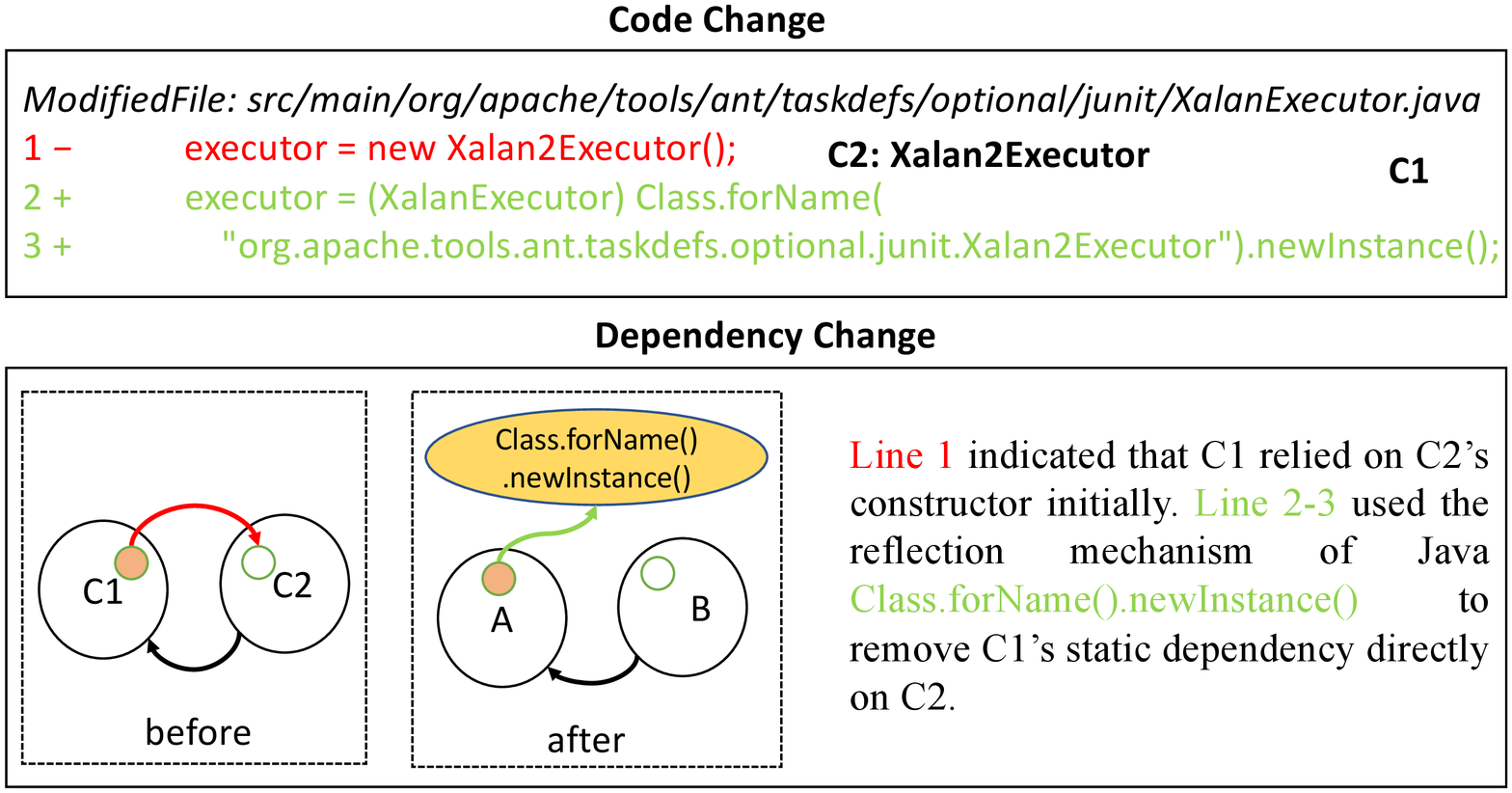}

	\caption{Leverage Built-In Feature}
	\label{fig:leverageBuiltIn}
\end{figure}

In Commit \#284e790\footnote{\url{https://github.com/apache/ant/commit/284e790}} of the Ant project, the commit message ``\textit{Make XalanExecutor independent of Xalan2 so one can compile Xalan1Executor without Xalan2}'' indicates that the purpose of this commit is to untangle the cycle between class \textit{XalanExecutor} and class \textit{Xalan2Executor}. The code changes are shown in Figure~\ref{fig:leverageBuiltIn}. We examined the code changes in class \textit{XalanExecutor} and discovered that, instead of instantiating class \textit{Xalan2Executor} directly using \textsc{new}, developers applied a built-in reflection method \textit{Class.forName().newInstance()}. This case leverages Java built-in reflection features to create an instance of class objects.

\subsubsection{Remove Unused/Deprecated Code} 

As shown in Figure~\ref{fig:pattern5}, when a feature becomes obsolete and is deprecated, the code which causes the dependency cycle can be removed from the code base. In our manual analysis, this is the most common case we encountered. We found that 40.0\%(235/587) of the two-class dependency cycles were resolved by simply eliminating the feature that caused the cyclic dependency.  

For example, in Commit \#829caa5\footnote{\url{https://github.com/apache/wicket/commit/829caa5}} of the Wicket project, one class, \textit{PageSet}, was removed, along with its reference code in its cycle class, \textit{Page}. We examined the commit message and the message stated that ``\textit{Experimental feature will not be supported anymore and really didn't have any use at it's current state}''. For another example, in Commit \#73819ca\footnote{\url{https://github.com/apache/ambari/commit/73819ca}} of the Ambari project, as developers ``\textit{completely removed admin role from ambari}'', the original dependency cycle between class \textit{RoleEntity} and class \textit{UserEntity} was also resolved as class \textit{RoleEntity} was removed.

\subsubsection{Associate with a Big Architectural Refactoring}
Except for the cases that we can categorize into the above five scenarios, there are several instances of dependency cycle resolution involving a significant amount of code addition and deletion at the architecture level. Identifying the patterns for these cases requires deeper domain knowledge, and consequently we classify them into the \textit{Associate with a Big Architectural Refactoring} category.

\begin{center}
\noindent\fbox{%
    \parbox{\linewidth}{%
        \textbf{To answer RQ1, our manual analysis and inspection across 38 open-source projects of different domains shows that developers tend to apply five recurring patterns in the untangling practice of two-class dependency cycles.} ``Remove Unused/Deprecated Code'' is the most frequent observed pattern, and ``Move Code From One Cyclic Class To a Third Class'', ``Move Code Between Two Cyclic Classes'', ``Shorten Call Chain'', ``Complex'', and ``Leverage Built-In Feature'' are following next. These recurring patterns indicate that developers frequently apply common practices to untangle dependency cycles. We believe that the taxonomy of these patterns can help developers understand dependency cycles better and thus tackle them more efficiently.}
    }%
\end{center}
\subsection{Design Characteristics in Dependency Cycles}
\label{design}

After we identified recurring patterns for untangling dependency cycles between two classes, one question raises up: are certain patterns applied to specific types of dependency cycles? In other words, we want to know if dependency cycles resolved using the same pattern exhibit similar characteristics. If the answer is positive, we can recommend that developers can use particular patterns to address dependency cycles with certain traits. If not, it suggests that the characteristics of dependency cycles may not be related to the chosen untangling methods, and developers may have selected these patterns at random. In this case, our study can also help the untangling process by providing all candidate patterns and potential fixes for developers to choose from.

Untangling dependency cycles involves changing the dependency relations between two cyclic files, sometimes dependency changes can even propagate to neighboring files. Intuitively, we examined whether dependency relations among cyclic files and their neighbors could determine the chosen untangling patterns. As various types of dependency relations may exist between two classes as shown in Table~\ref{tbl:dp_types}, we investigated whether and how different types of dependency relations within cyclic files, and among cyclic files and neighboring files, affect the chosen patterns. We try to solve the problem whether the design characteristics inside and outside of dependency cycles can determine the chosen untangling pattern. Basically it is a classification problem. Our initial thought was to embed dependency relations within and outside a cycle into vectors and check if a trained Graph Neural Network (GNN) model could classify them into different categories. However, as far as we know, current GNN models do not have a mature method for dependency calling graph's representation, and such graph representation learning also requires a large amount of data~\citep{xu:2021understanding}. Therefore, in this study we represented different types of dependency relations as features and adopt classic machine learning methods to examine the relations between dependency relation traits and different recurring patterns of cycles' untangling.

\subsubsection{Design Characteristics within Cycles}

We first investigated internal design characteristics inside two cyclic files. Since dependency graphs are bidirectional, in order to fairly compare them in different cycles, we first set certain dependency relations as dominant, labeling the file initiating such relations as f1 and the other file in the cycle as f2. In this way, we labeled all dependency relations as ``depType\_fi\_fj'', which means fi depends on fj with depType. In our experiment, if there exists inheritance relations denoted as ``Extend'' or ``Implement'', then we mark it as an dominant relation. If no inheritance relations exist, then we mark ``Call'' as dominant. For example, among 5 untangling cases in the ``Leverage Built-In Features'' category in Table~\ref{tbl:builtin}, 3 cases have an ``Extend'' relation; we labeled the file initiating this relation as f1 and considered the other cyclic file as f2. If an ``Extend"  relation is already marked as ``f1\_f2'' and another ``Call'' relation exists between f2 and f1, it is marked as ``Call\_f2\_f1''. After accurately marking these dependency relations, one observation is that the most frequent relations in the ``Leverage Built-In Features'' pattern involve f1 extending f2 while f2 creates f1. This relation accounts for 50\% of all cases (cases 1, 2, and 3). The second most frequent relation is f1 calling f2 while f2 uses f1, accounting for the remaining 50\% of all cases (cases 4, 5 and 6).

\begin{table}[!ht]
	\caption{Dependencies in Leverage Built-In Feature}
	\centering
	\label{tbl:builtin}
\begin{tabular}{|c|l|c|}
\hline
1 & \textcolor{red}{Extend\_f1\_f2}, \textcolor{blue}{Create\_f2\_f1}, Call\_f1\_f2, Use\_f1\_f2 & \multirow{3}{*}{50\%}\\ \cline{1-2}
2 & \textcolor{red}{Extend\_f1\_f2}, \textcolor{blue}{Create\_f2\_f1}, Call\_f1\_f2, Call\_f2\_f1 & \\ \cline{1-2} 
3 & \textcolor{red}{Extend\_f1\_f2}, \textcolor{blue}{Create\_f2\_f1}, Call\_f2\_f1 & \\ \hline 
4 & \textcolor{red}{Call\_f1\_f2}, \textcolor{blue}{Use\_f2\_f1}, Create\_f1\_f2 &\multirow{3}{*}{50\%} \\  \cline{1-2}
5 & \textcolor{red}{Call\_f1\_f2}, \textcolor{blue}{Use\_f2\_f1}, Create\_f1\_f2 & \\  \cline{1-2}
6 & \textcolor{red}{Call\_f1\_f2}, \textcolor{blue}{Use\_f2\_f1}, Cast\_f1\_f2 & \\ \hline 
\end{tabular}
\end{table}

Using the above method, we annotated all types of dependency relations and 24 unique types of dependency relations were extracted from all 536 cases with five untangling patterns. The details of these 24 unique types can be obtained from our replication package~\citep{Replication}. For each unique ``depType\_fi\_fj'', if a cycle contains it, then we mark it as 1, otherwise 0. In this way, we represented each cycle with one 24-length array with each feature's value marked as 1 or 0. Next, we applied classic machine learning methods to see if these cases can be classified into their untangling patterns. Since the instances of these five patterns vary significantly, we used a technique called SMOTE \citep{chawla:2002smote} to get a balanced dataset by over sampling and creating synthetic minority class samples. After that, we split the dataset to 80\% of training and 20\% of testing data. Last, we trained ``Nearest Neighbors''(n\_neighbors=5), ``Linear SVM''(kernel="linear", C=0.025), ``Decision Tree'', (max\_depth=10), and ``Multi-layer Perceptron''(MLP, alpha=1, max\_iter=1000) with the default setting from the scikit-learn package with the training data and tested the model with the testing data. We found that, among these classifiers, ``MLP'' achieved the highest average accuracy in all untangling patterns.


\begin{table}[th]
	\caption{MLP Classification Using Dependencies Inside Cycles}
	\centering
	\label{tbl:depInCycle}
\begin{tabular}{|l|r|r|r|}
\hline
\textbf{Untangling Patterns} & \textbf{Precision} & \textbf{Recall} & \textbf{F1-score} \\ \hline 
Move Code from One Cyclic File to a Third Class & 0.68 & 0.42 & 0.52 \\ \hline 
Move Code between Two Cyclic Files & \textcolor{red}{0.59} & \textcolor{red}{0.64} & \textcolor{red}{0.61} \\ \hline 
Remove Unused/Deprecated Code & 0.38 & 0.48 & 0.42 \\ \hline 
Shorten Call Chain & 0.62 & 0.58 & 0.60 \\ \hline 
Leverage Built-in Feature & \textcolor{red}{0.75} & \textcolor{red}{0.98} & \textcolor{red}{0.85} \\ \hline 
\end{tabular}
\end{table}

Table~\ref{tbl:depInCycle} shows MLP's prediction details. From this table, we can see that ``MLP'' can predict cycles which use the ``Leverage Built-In Feature'' pattern with a precision 0.75 and a recall 0.98. This results suggest that the internal dependency structure inside this pattern already shows distinct characteristics, and consequently ``MLP'' can use its distinct dependency relations to predict the pattern the dependency cycle can use. This is not surprising, as our previous analysis (see Section~\ref{patterns}) revealed that 50\% of this pattern involves f1 extending f2 while f2 creates f1. Upon manual inspections, we found that this type of dependency relations only exist and are unique in this pattern. The second best result of MLP's prediction is ``Move Code between Two Cyclic Files'' with F1-score 0.61. We also manually inspected these cases and found that ``Use\_f2\_f1'' and ``Call\_f1\_f2'' are exclusive to the ``Move Code between Two Cyclic Files'' pattern. These unique traits inside the dependency cycle provide the reason why why ``MLP'' can classify these patterns with a relatively high precision and recall.

Though MLP can successfully predict ``Leverage Built-In Feature'' and ``Move Code between Two Cyclic Files'' based on dependency cycles' internal dependency traits, we also observed that, the F1-score of ``Shorten Call Chain'', ``Move Code from One Cyclic File to a Third Class'', and  ``Remove Unused/Deprecated Code'' is 0.60, 0.52, and 0.42 separately. It seems that ``MLP'' classifier cannot tell which one of these three patterns to use if only based on dependency cycles' internal structure. It also suggests that if such a cyclic case occurs, ``MLP'' classifier can likely choose any of these three untangling strategies. Different from ``Leverage Built-In Feature'' and ``Move Code between Two Cyclic Files'' whose internal dependency relations already show specific traits and untangling pattern only involves cyclic classes, ``Move Code from One Cyclic Class to a Third Class'' and ``Shorten Call Chain'' also involve a third class's participation. Thus, we conjecture that the neighbor files' interaction with the dependency cycle can play an important role in deciding which pattern to use.

\subsubsection{Design Characteristics outside Cycles}

To verify the above assumption, next we investigated design context outside two cyclic classes. We tracked all neighbor files, which have direct relations and co-changed with cyclic classes, and extracted their dependency relations. Similarly, we distinguished the dominant file f1 and the other file f2. Then, we marked all neighbor files as f3 and used the same method to annotate the dependency relations between f3 and f1/f2. For example, if f1 ``extends'' f3, then we annotate it as ``extend\_f1\_f3''; if a neighbor file ``uses'' f2, then we annotate it as ``use\_f3\_f2''. In this way, we extracted another 52 unique dependency relations between neighbor and cyclic files, and represented a dependency cycle' design context as a 52-length array. Different from two cyclic classes, it may exist more than one neighbor files with the same types of dependency with cyclic files. We also added weights to these 52 unique dependency relations. For example, if there exists 3 neighbor files ``call'' f1, we represent this ``call\_f3\_f1'' feature's value as 3. We combined these 52 features of the cycle's design context with the cycle's internal 24 features to form a 76-length array. Finally, to better compare the classification effect using only cyclic internal dependency relations, we also adopted SMOTE to create a balanced dataset and trained the ``MLP'' model with 80\% of data with 76 features. Finally, we used the trained ``MLP'' to classify 20\% of testing data, and the result is shown in Table~\ref{tbl:depOutCycle}.


\begin{table}[th]
	\caption{MLP Classification Using Dependencies Inside and Outside Cycles}
	\centering
	\label{tbl:depOutCycle}
\begin{tabular}{|l|r|r|r|}
\hline
\textbf{Untangling Pattern} & \textbf{Precision} & \textbf{Recall} & \textbf{F1-score} \\ \hline 
Move Code from One Cyclic File to a Third Class & 0.71 & 0.59 & 0.64 \\ \hline
Move Code between Two Cyclic Files & 0.64 & 0.63 & 0.64 \\ \hline
Remove Unused/Deprecated Code & 0.52 & 0.57 & 0.54 \\ \hline
Shorten Call Chain & 0.67 & 0.59 & 0.63 \\ \hline
Leverage Built-in Feature & 0.76 & 0.98 & 0.86 \\ \hline
\end{tabular}
\end{table}

From Table~\ref{tbl:depOutCycle}, we can see that the precision is slightly improved for ``Leverage Built-In Feature'' (0.85$\rightarrow$0.86) and ``Move Code Between Two Cyclic Classes'' (0.61$\rightarrow$0.64). It means that the design context outside dependency cycle can help developers choose these two patterns better. Furthermore, we observed the precision and recall is significantly improved for ``Move Code from One Cyclic Class to a Third Class'', with precision from 0.68 to 0.71, recall from 0.42 to 0.59, and F1-score from 0.52 to 0.64. This suggests that the dependency relations between neighbor files and the dependency cycle play an important role for the cycle' untangling pattern. It makes sense as we mentioned earlier that this pattern involves a third class's participation so the neighbor files' dependency relations with the cycle is definitely very important. We also observed F1-score is slightly improved for the ``Shorten Call Chain'' pattern and ``Remove Unused/Deprecated'' pattern.

\begin{center}
\noindent\fbox{%
    \parbox{\linewidth}{%
        \textbf{To answer RQ2, ``MLP'' prediction results show that recurring patterns do exhibit specific characteristic signs.} First, the patterns developers chose in practice to untangle dependency cycles are highly related with cycles' internal dependency relations. Especially for ``Leverage Built-In Feature'' and ``Move Code between Two Cyclic Classes'', the dependency relations inside the cycle can determine the applied pattern. Second, we found that, combining the design context outside a dependency cycle with the dependency relations inside the cycle can have a better prediction of untangling patterns, especially when the pattern involves a third class's participation.  
    }%
}
\end{center}
\subsection{Counterintuitive Solutions in Untangling Dependency Cycles}
\label{mistakes}

Not only recurring patterns were applied to effectively untangle cycles, in our empirical study we also observed 69 dependency cycles were not correctly addressed. These cases usually resulted in breaking the original cycle, only to form a new, sometimes even larger, cycle. While it is possible that the intention of that particular commit is not to untangle the cycle, we argue that developers also do not intend to create new or larger cycles. However, there may be complex reasons such as trade-offs or deliberate design choices behind these cases. Therefore, we refer to these cases as `counterintuitive solutions' in the paper. Our manual analysis shows three common counterintuitive solutions developers always used in addressing these dependency cycles. Revealing and understanding these common counterintuitive solutions can help developers avoid similar errors in the future. In this section, we numerate the three common counterintuitive solutions observed in our empirical study and discuss how to avoid them.

\subsubsection{Cycle Shift to a Parent Class}

In this scenario, a child class and a third class originally are calling each other, forming a cycle. Meanwhile, the third class also depends on the parent class. During a commit, the child class \emph{moves up} a method, which depends on the third class, to its parent class. As a result, the cycle shifts from between the child and the third class to between the parent and the third class. 

\begin{figure}[!thbp]
	\centering
	\includegraphics[angle=90,width=\linewidth,trim={5cm 1cm 4cm 1cm}]{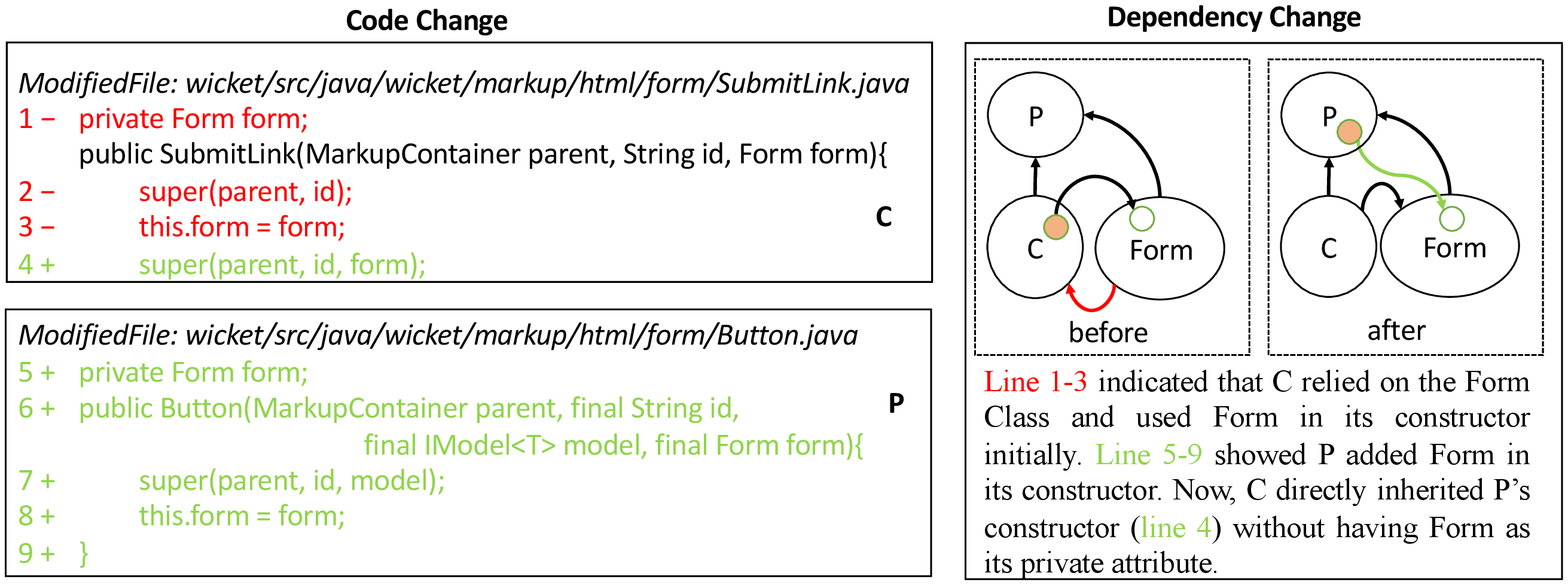}
	\caption{Cycle Shift to a Parent Class}
	\label{fig:cycleshift2parent}
\end{figure}

Commit \#f96f99e\footnote{\url{https://github.com/apache/wicket/commit/f96f99e}} of the Wicket project in Figure~\ref{fig:cycleshift2parent} shows such an example. A child class \textit{SubmitLink} (denoted as C) \emph{moved up} its constructor, in which class \textit{Form} is a parameter, to its parent class \textit{Button} (denoted as P). The commit message ``\textit{AjaxSubmitButton and AjaxSubmitLink now extends Button. This gives them the possibility to set DefaultFormProcessing}'' denoted that the reason for this change was to enable the parent class \textit{Button} to pass the parent's constructor to more children classes. Meanwhile, the changes in this commit indicated the third class \textit{Form} did not call the child class \textit{SubmitLink} directly any more. Consequently, the cycle shifts from between the child C and the third class \textit{Form} to between the parent P and the third class \textit{Form}. Though manual examination, we found that the constructor, which was moved up from this child to parent class, used the entire third class \textit{Form} as a parameter. Since \textit{Form} and \textit{Button} are UI components, we can apply ``\textit{inversion of control}'' to decouple these two classes and resolve this cycle. For example, we can assign a callback to the clickable event of the button and implement an observer pattern for \textit{Form}, \textit{Button} and the event class. As this involves more architectural change, it would belong to \emph{Associate with a Big Architectural Refactoring} pattern mentioned in Section~\ref{patterns}.

\subsubsection{Cycle Shift to a Child Class}

In this scenario, a parent class and a third class depend on each other, forming a cycle, and a child class also depends on the third class. In a commit, the third class removed the dependency to the parent class, and instead called the child class, causing the cycle to shift from between the parent and the third class to between the child classes and the third class. 

\begin{figure}[!thbp]
	\centering
	\includegraphics[angle=90,width=\linewidth,trim={4cm 2cm 3cm 2cm}]{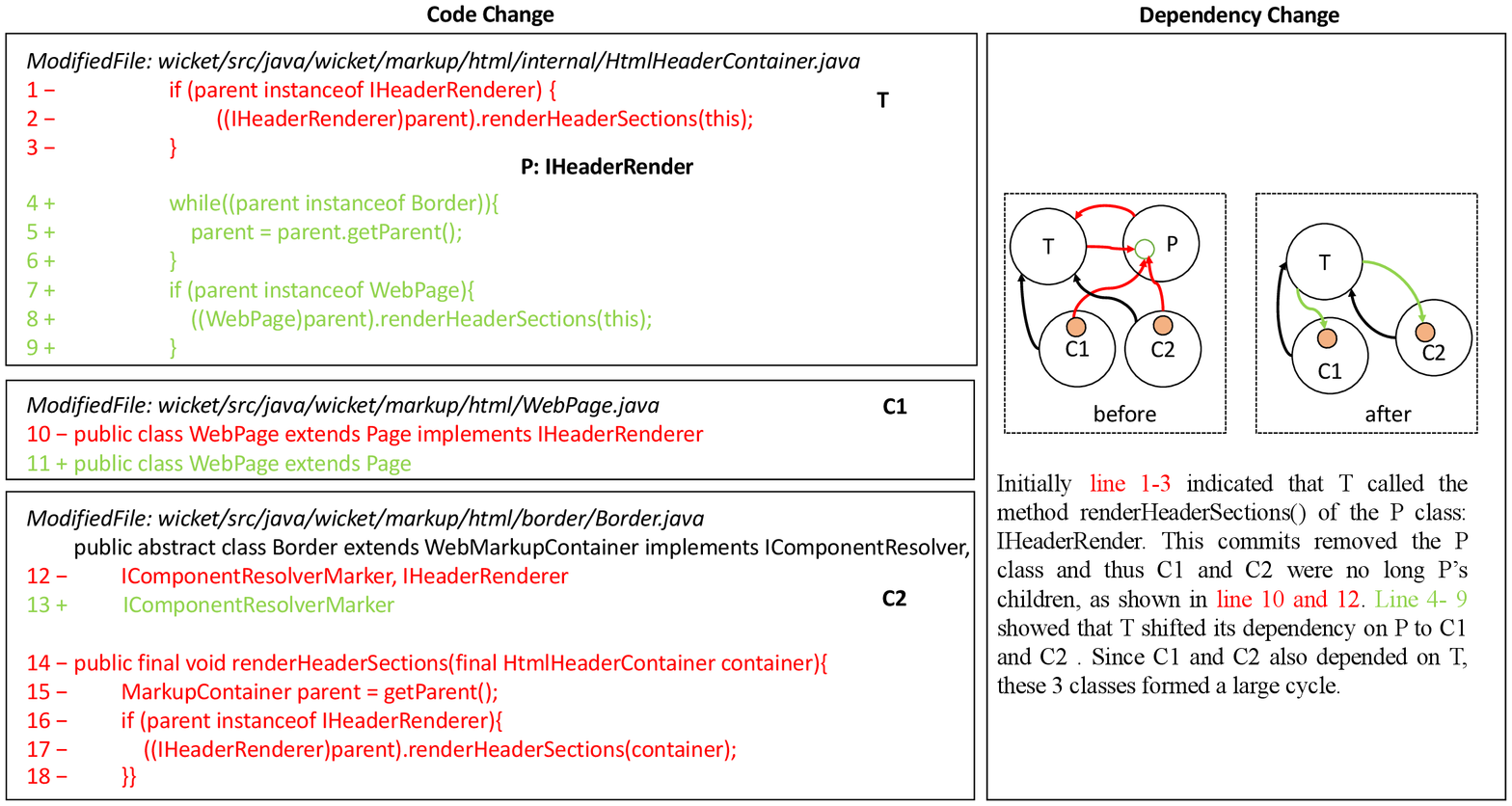}
        \centering
	\caption{Cycle Shift to Children Classes}
	\label{fig:cycleshift2child}
\end{figure}

Figure~\ref{fig:cycleshift2child} shows the modified code in the~\textit{HtmlHeaderContainer} class of Commit \#763dcc3\footnote{\url{https://github.com/apache/wicket/commit/763dcc3}} in the Wicket project. Line 1-9 show that class \textit{HtmlHeaderContainer} (denoted as T) deleted the code which calls the parent class \textit{IHeaderRenderer} (denoted as P), and added dependencies on its original children classes: \textit{WebPage} (denoted as C1) and \textit{Border} (denoted as C2). However, the cycle is not resolved. The dependency change shows that after the commit, the cycle shifts from between P and T to a larger cycle involving T, C1, and C2.

Upon further examination, we found that class P got deleted in this commit. Subsequently, class C1 and C2 removed its implementation to class P and the overridden method ``renderHeaderSections'' from class P also got deleted, as shown in Line 10-11 and Line 12-18. However, we found that class C1 and C2 both depended on class T. Class C1 ``imports'' class T. Class C2 also overrode a method of class \textit{WebMarkupContainer} which uses class T as a parameter. This results in two cyclic dependency: T and C1, T and C2, forming a larger dependency cycle with 3 classes. To avoid this problem, when shifting a third class's dependency from parent to children classes, developers first need to examine whether children classes have any dependencies to the third class and whether these dependencies can be removed. Though the unnecessary ``import'' from C1$\rightarrow$T can be removed, class C2's dependency on the third class T is deeper and cannot be removed directly. In this case, simply shifting dependencies cannot untangle the cycle.

\subsubsection{Cycle Shift to a Third Class}

\begin{figure}[!thbp]
	\centering
	\includegraphics[angle=90,width=\linewidth,trim={6cm 1cm 6cm 1cm}]{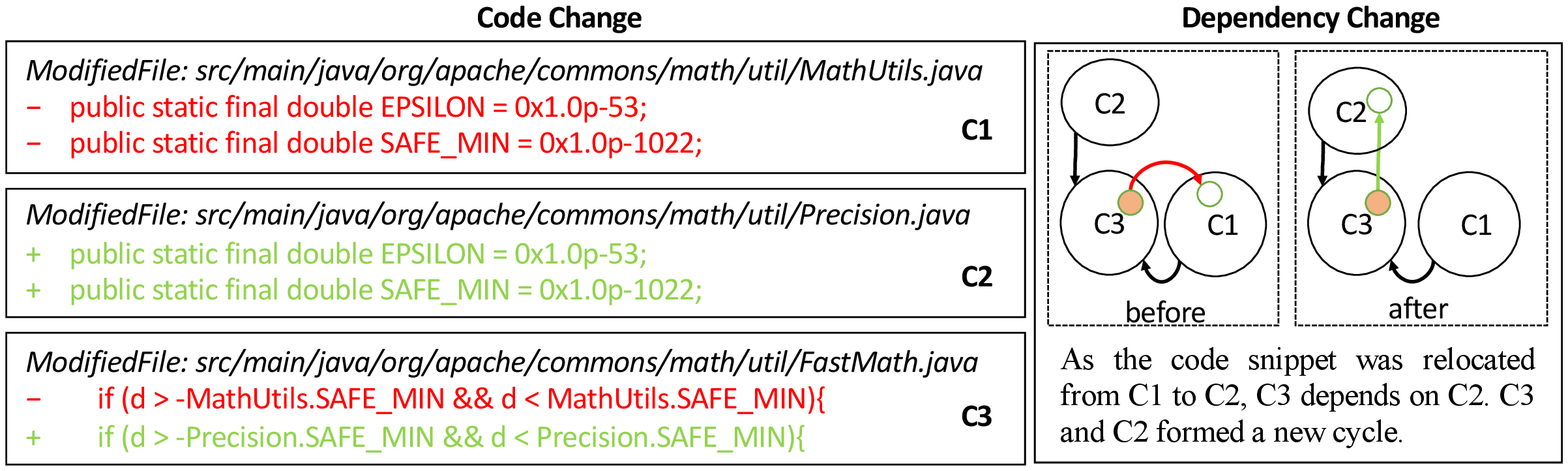}
        \centering
	\caption{Cycle Shift to a Third Class}
	\label{fig:cycleshift2third}
\end{figure}

Compared to the above two scenarios happening in parent and child classes, this scenario is more likely to occur in utility features. In Commit \#3a7d733\footnote{\url{https://github.com/apache/commons-math/commit/3a7d733}} of the Commons-math project shown in Figure~\ref{fig:cycleshift2third}, class \textit{FastMath} (denoted as C3) originally used the static attributes in the \textit{MathUtils} class (denoted as C1). This commit was associated with the issue MATH-689\footnote{\url{https://issues.apache.org/jira/browse/MATH-689}} in the Jira issue tracking system, whose intent is to break up the MathUtils class. 

As two static attributes in class C1 were relocated to the Precision class (denoted as C2), class C3 now relied on class C2. All of these three classes were placed in the same ``util'' package. However, both class C1 and class C2 also depended on class C3. Consequently, the original dependency cycle between class C1 and class C3 was shifted to a new cycle between class C2 and class C3 after the static attributes' relocation. Upon further examination, we discovered that the static attributes were not used in either class C1 or C2. Instead, these attributes are more closely coupled with class C3 than with either class C1 or C2. So moving these attributes from class C1 to C2 is to  ``\textit{move attributes from one wrong place to another}''. The optimal solution to break the cycle is to move the static attribute to class \textit{FastMath} where the static attributes are more closely coupled or create a new class for storing these static attributes.

\begin{center}
\noindent\fbox{%
    \parbox{\linewidth}{%
        \textbf{To answer RQ3, among all cases involving cycles' addressing, we found that 10.5\% (69/(69+587)) of dependency cycles were not correctly untangled.} These cases show some common characteristics and can be classified into the three categories we summarized above. We believe that the awareness of these counterintuitive solutions can help developers better handle cycle's untangling cases.
    }%
}
\end{center}

\section{Discussion}
\label{sec:diss}

In this section, we first present the analysis of the results and discuss the implications of our findings, and then we clarify the threats to the validity of our analysis. 

\subsection{Analysis of Results}
\textbf{RQ1}: Through hundreds of dependency cycle untangling cases from different projects in various domains, we have found in RQ1 that developers tend to apply five recurring patterns to untangle dependency cycles between two classes in practice. These five recurring patterns account for 94.8\% of all untangling cases, which indicates that these patterns are frequently used by developers and prevalent in the code repository. These summarized patterns can be shared with practitioners to improve their understanding of effective strategies for resolving two-class dependency cycles. Furthermore, the findings of our study can be incorporated into educational materials and training programs for software developers, improving their understanding of dependency cycles and how to address them.
 
\textbf{RQ2}: We demonstrated in RQ2 that, based on cycles' internal structure and design context, machine learning models can be trained to predict the untangling pattern of two-class dependency cycles. This finding suggests that, similar to the state-of-the-art automated program repair techniques, which were achieved by training code embedding ~\citep{li:2020icse, jiang:2021icse}, it is possible to leverage machine/deep learning techniques in assisting dependency cycles' refactoring and setting up cycles' refactoring goals/patterns. It is worth to mention that some patterns do occur more frequently in some particular projects, but we had not observed that a single project involves only one pattern. We will investigate whether projects from the same domain may achieve a better prediction result in the future work.

\textbf{RQ3}: We have revealed three common counterintuitive patterns of addressing these two-class dependency cycles in RQ3. Among all cases of addressing cycles, we found that 10.5\% (about one in ten) of dependency cycles were not correctly addressed and fell into these three counterintuitive solutions. While conducting follow-up interviews with developers about their decision-making process could provide better insights into why they chose these counterintuitive solutions, we have also observed some simple mistakes directly from code changes, such as the case illustrated in Figure~\ref{fig:cycleshift2third}. It means that dependency cycles' handling is challenging and error-prone.  This result is consistent with other refactoring empirical studies by \cite{bavota:2013icse}, \cite{kim:2014tse}, and \cite{sharma:2015software}, in which they claimed that architectural refactoring is risky and needs strong support and guidance. In Section~\ref{mistakes}, we discussed how the three counterintuitive solutions of addressing two-class dependency cycles can be better handled. We hope that this information can be used to develop guidelines and best practices for practitioners, helping them avoid repeating these counterintuitive solutions. 

\subsection{Implications}

\textbf{A simple dependency cycle's untangling is not trivial.} \cite{fowler:1999book} recommended to apply the camp site rule---\textit{``always leave the code behind in a better state than you found it''}. By doing this, small regular refactoring can be combined in developers' daily activities, such as implementing new features, fixing bugs, and improving code base health. Significant advancements have been made in the field of detecting local refactoring opportunities through various research~\citep{terra:2018jmove, tsantalis2009identification, cui:2022rmove, liu2015identifying}. Meanwhile, refactoring operations (``move method'', ``pull members up'', etc.) are supported in some popular IDEs, such as IntelliJ and Eclipse. While these techniques provide strong support for code refactoring, refactoring at the design or architectural level still lacks valid support, especially for legacy software systems~\citep{kim:2014tse,sharma:2015software}. Our study results show that, with all these support, a significant ratio of mistakes still occur in addressing a simple two-class dependency cycles. Also, there exist five different patterns to resolve a simple two-class dependency cycles. This evidence shows that even a simple dependency cycle's untangling is not trivial. Refactoring, especially that involving multiple files, is challenging for software practitioners~\citep{lacerda:2020jss,peruma:2022refactor,bavota:2013icse} and needs better support.

\textbf{A top-down refactoring approach can be benefited by leveraging internal design characteristics of two-class dependency cycles and their design context.} For complex refactoring at the design or architectural level, architects tend to adopt a top-down approach, which is to set a refactoring goal first and then design detailed steps towards the goal~\citep{lin:2016interactive}. However, the settings of a refactoring goal heavily reply on developers' experience and domain knowledge and cannot be automated currently. Our study shows that the untangling patterns (or refactoring goals) of two-class dependency cycles can be predicted by leveraging the internal design characteristics of two-class dependency cycles and their design context. This finding suggests that, similar to program repair which can be automated by learning code relations~\citep{li:2020icse, jiang:2021icse}, dependency cycles' repair/refactoring goals can also be semi-automated theoretically. The challenge is how we can mine the ``hidden'' knowledge and patterns at the architectural or design level. Furthermore, two important questions raise up: (1) how to represent the information of nodes and edges in and out of dependency cycles with more than 2 files; and (2) how to train machine learning or deep learning models to better leverage dependency cycles' internal and context information. The answers to these questions are valuable for automating refactoring goals of complex dependency cycles or other architecture anti-patterns~\citep{mo:2019tse}. We believe that the answers to these questions are the keys towards better refactoring support at the architecture or design level.

\subsection{Threats to Validity}

\textbf{Construct Validity}: The analysis of the recurring patterns and common counterintuitive solutions of cycles' untangling heavily relies on the manual inspection. In order to reduce human bias and neglect, we designed a protocol to track code snippets' changes in commits and summarize patterns. And four experienced developers divided into two groups participated in the inspection process, until all developers reached an agreement. Besides, we also cross-validated the identified patterns from our manual inspection with \emph{RefactoringMiner}~\citep{tsantalis:2020tse}, which is a refactoring mining tool and can mine atomic refactoring operations from code changes. For example, if through manual inspection, all developers agreed that it is a ``move code between two cyclic classes'' pattern, we used \emph{RefactoringMiner} to verify that a ``move method/attribute/static class'' operation was also detected. We also used \emph{DV8}\footnote{\url{https://archdia.com}} (a dependency graph visualization tool) to verify the dependency change if all developers agreed that it is a ``move code from one cyclic class to a third class'' or ``shorten call chain'' pattern.  

\textbf{Internal Validity}: One major internal threat is the way we represent dependency relations in and outside of cycles. There exist many ways for dependency graph representation. In this work, we chose ``depType\_fi\_fj'' as it is most straightforward to present nodes and directed edges of dependency graphs, and contains necessary structure information of cycles and their context. Also, the high precision and F1-scores using the ``MLP'' model to classify untangling patterns demonstrate its effectiveness. We are not sure whether different representations of cycles~\citep{cai:2018tkde}, such as deep walk~\citep{perozzi:2014deepwalk}, node2vec~\citep{grover:2016node2vec}, etc., will produce a different result. We leave the comparison of prediction refactoring patterns with different representations as our future work.

\textbf{External Validity}: In terms of external threats, we only studied 263 untangling cases in Java in our study since it involves intensive manual inspections. It remains unclear whether our findings can be generalized to other open-source projects or closed-source industrial projects. Though in this study, 18 projects were selected from different domains and can be representative, we cannot guarantee that these observed patterns and common counterintuitive solutions are universal. We plan to validate this by analyzing more diverse projects in different programming languages in the next step.

\textbf{Conclusion Validity}: Our approach detected dependency cycles' fix and summarized the fix patterns in a single commit. However, it is possible that developers may take multiple commits to fix a dependency cycle. In such a situation, the untangling patterns of the whole process with multiple commits may be different. In this work, as we focus on the last step of two-class dependency cycles' untangling, the conclusion of the summarized patterns is valid if we narrow down the scope. Moreover, in order to increase the reliability of the results, we made the replication package of this work available online~\citep{Replication}.

\section{Related Work}
\label{sec:related}

The study of dependency cycles has been a topic of interest in the software engineering research community for many years. In this section, we discuss the related research of our work from two perspectives: empirical studies on dependency cycles, which focus on the causes and consequences of dependency cycles, and refactoring of dependency cycles. 

\subsection{Empirical Studies of Dependency Cycles}
Numerous empirical studies have been conducted to understand dependency cycles and their impact on software systems. \cite{dietrich:2010barriers} and~\cite{melton:2007empirical} showed that dependency cycles are pervasive in modern software systems, affecting code's comprehension and testing. \cite{zazworka:sqj13} investigated the impact of dependency cycles on software maintainability, and the results show that that systems with a higher number of dependency cycles have lower maintainability scores. \cite{maccormack:mnsc06} performed a case study on a large-scale software system and found that dependency cycles were often caused by architectural erosion~\citep{Li2022SMS}, which led to increased complexity and reduced modularity.~\cite{lu:2016icse} defined a particular dependency cycle model called Hub, in which a center file depends on a set of other files and that set of files also depend on the center file. They analyzed the maintenance effort of Hub through different releases and concluded that Hubs have been growing in size through releases and cost large maintenance effort in terms of defects and changes. \cite{snipes:2018seaa} conducted a case study about the effects of architecture debt on software evolution effort. They proved that files involved in dependency cycles are highly correlated with the maintenance efforts. \cite{oyetoyan:2013jss} classified software components into two groups – the cyclic and the non-cyclic ones, and their results show that most defects and defective components are concentrated in cyclic-dependent components, either directly or indirectly.

Recently, more and more empirical research focuses on studying how dependency cycles get evolved in the code revision.~\cite{oyetoyan:2014csmr} analyzed the evolution of dependency cycles among components. By studying dependency cycles through different releases of software systems, they found that there is no evidence of any systematic “cycle-breaking” refactoring in these dependency cycles among components. A recent study by~\cite{feng:2023jsep} explored how dependency cycles among classes evolve at the commit level. Their results show shows that dependency cycles with different topological structure can present different evolution characteristics. While the above empirical studies provide valuable input about dependency cycles' evolution and their impacts on software quality, how dependency cycles get resolved by practitioners is still not clear. In this work, we try to fill this gap and study how code snippets are removed or replaced in addressing two-class dependency cycles. Through the manual inspection with a well-defined study protocol, our study  identified five recurring patterns and three common mistakes in dependency cycles' untangling process. Besides, we also prove that the untangling patterns chosen by developers are not only determined by the internal structure of dependency cycles, but also highly related to the design context of the dependency cycles.

\subsection{Refactoring of Dependency Cycles}

Several studies have proposed effective strategies and best practices for (semi-)automatically refactoring dependency cycles~\citep{shah:2012ecsme,shah:2013icsm,goldstein:2014icse,oyetoyan:2015icsme,caracciolo:2016saner}. \cite{shah:2012ecsme} introduced an algorithm that eliminates circular dependencies between packages by relocating classes between them. In addition to moving classes, they suggested refactoring operations such as type generalization and service locator to resolve circular dependencies. The effectiveness of their approach was validated using instances of cyclic dependencies, demonstrating a decrease in their occurrence after applying the proposed method.~\cite{goldstein:2014icse} introduced a method for automatically untangling cyclic dependencies among components. Their algorithm aims to minimize the number of classes to be relocated while simultaneously considering architectural metrics. They asserted that their approach not only resolved dependency cycles, but also improved architectural metrics like cohesion and coupling. 

\cite{oyetoyan:2015icsme} introduced a novel metric to evaluate coupling changes between a class's CRSS (the Class Reachability Set Size) and its interfaces, and developed a decision support system for breaking dependency cycles using this metric. Their assessment indicated that this new metric can help identify a smaller number of candidate classes for resolving large dependency cycles, ultimately decreasing the refactoring effort required.~\cite{caracciolo:2016saner} explored various refactoring strategies, recommending the most cost-effective sequence of operations to break dependency cycles. They determined the optimal strategy using a profit function, and concluded that their approach successfully eliminates cyclic dependencies between packages.~\cite{ferreira:2023tse} claimed that the orderings of recommended refactoring is difficult for developers to understand. They proposed an algorithm for detecting these dependencies among refactoring operations and defined refactoring recommendations as sets of refactoring graphs instead of sequences.

As a complement to existing research, we studied how developers addressed two-class dependency cycles in practice. As untangling cyclic dependencies is an NP-hard problem~\citep{goldstein:2014icse}, there may exist various ways to untangle a dependency cycle. Our study identified five recurring patterns in untangling two-class dependency cycles. We believe that these patterns have the potential to help developers towards automatic untangling of such dependency cycles in practice.

\section{Conclusions}
\label{sec:conclude}

In this paper, we conducted an empirical study on how dependency cycles between two classes get resolved in practice from 38 open-source projects, while maintaining the original functionalities. Our results show that developers tend to apply five recurring patterns to untangle two-class cycles. These patterns can be observed in projects from different domains. Moreover, developers also make common counterintuitive solutions in addressing dependency cycles. Our study can serve as a taxonomy to improve developers' awareness for dependency cycles' refactoring and also be used as learning materials for students in software engineering and inexperienced developers. 

To verify whether the design characteristics can determine the chosen untangling pattern, we extracted fine-grained dependency relations inside and outside dependency cycles as features and check whether these features can be used to classify the patterns. We have showed that for cycles' refactoring which only needs code changes inside cyclic classes, using features of the internal structure in cycles can achieve a good result of predicting the chosen pattern. But when a cycle' refactoring requires a third class's participation, the dependency relations outside dependency cycles need to be taken in consideration. This empirical study shows that it is theoretically feasible to use design characteristics to predict the refactoring goal of a simple dependency cycle. However, it is unknown whether it can be applied to more general cases. 

We plan to study whether a good prediction can also be achieved for complex dependency cycles in our future work. In that case, refactoring goals can be automatically set based on a large amount of learning data. Moreover, similar to JDeodorant \citep{tsantalis:2018saner} which is an Eclipse plugin and supports the refactoring of five typical code smells, we plan to implement an IDE-plugin to better support the detection and refactoring of two-class dependency cycles. 
\section*{Acknowledgements}
This work is supported by the National Natural Science Foundation of China (NSFC) under Grant No. 62172311.

\section*{Data Availability Statements}
The data generated and analyzed during the current study is available in the Zenodo repository at~\citep{Replication}.

\bibliographystyle{spbasic}
\bibliography{refs.bib}

\end{document}